\documentstyle[11pt,pasp,twoside,epsf]{article}
\def\edcomment#1{\iffalse\marginpar{\raggedright\sl#1\/}\else\relax\fi}
\reversemarginpar

\begin{document}
\title{Populations of Transient Galactic Bulge X-Ray Sources}
 \author{Jean Swank and Craig Markwardt}
\affil{Code 662, Goddard Space Flight Center, Greenbelt, MD 20771, U.S.A.}
%\author{Craig Markwardt}
%\affil{Code 662, Goddard Space Flight Center, Greenbelt, MD 20771, U.S.A.}

\begin{abstract}
Starting in 1999, the Rossi X-ray Timing Explorer (RXTE) has monitored
the central bulge region of the Galaxy with the Proportional Counter 
Array (PCA), resolving 
about 50 binary X-ray
sources, including 18 sources discovered by RXTE and BeppoSAX. The
accretion rates that RXTE observed from these sources ranged from 
highs approaching Eddington limits to lows that may
correspond to mass exchange for a binary period near the minimum of 80
minutes. Several neutron star binaries with low peak luminosity have
outburst or cycle time-scales which are shorter than those of brighter
and better known counterparts.  We compare the characteristics of the
binaries with low rates of mass exchange to predictions of their
evolution.

\end{abstract}

\section{Introduction}

The bright Galactic bulge populations were among the first known
compact X-ray sources. The neutron star binaries are the ``Z sources''
and the ``Atoll sources''.
% (Hassinger \& van der Klis 1989). 
While many sources of Type I bursts (thermonuclear flashes) are Atoll
sources, there are others that have low persistent fluxes 
which have not been well studied.
Of course there are also the 
transient black hole candidate sources such as Nova Ophiuchus and 
XTE J1748-280. 

Transients have been seen by BeppoSAX and RXTE which have peak
luminosities of 100 mCrab rather than 1-10 Crab. Some are probably
intrinsically bright, but situated on the other side of the
galaxy. For others however there is distance or other evidence that
they are of lower luminosity.  

A famous low luminosity neutron star source is SAX~J1808.4--3658,
the transient burster that is a coherent 2.5 msec pulsar in a 2 hour
binary (Chakrabarty \& Morgan 1998). The BeppoSAX Wide Field Camera (WFC)
has identified
17 new burst sources (in't Zand 2000, Heise et al. 1999, Sidoli et al. 1999).
Heise et al.
suggested that a new class of sources with low peak luminosities
was present in the bulge region. King (2000) pointed out that evolution
of low-mass neutron star binaries would be expected to lead to small
systems with low peak luminosities and short time-scales similar to
those of SAX~J1808.4--3658. 

Among low luminosity black hole candidates, GRS~1737-31 
%(Cui et al. 1997),
had a spectrum similar to that of Cyg~X-1 in the low state and
aperiodic variability producing a similar percentage variability of
$\approx 30 \%$. These characteristics provide less secure
identifications than Type I bursts provide
for neutron star sources. 

Some sources have been present throughout the observing period and are
long-lived, some variable, some steady.  The sources which have fluxes
less than a mCrab may be bright for Chandra and XMM-Newton and more 
sensitive 
observations and identifications  will develop a more complete picture.

\section{RXTE Observations}

%\begin{figure}
%\plotfiddle{xscan-1999-06-14.ps}{3.5 in}{90.}{50}{50}{180}{0}
%\caption{Sample east-west scan path. Dashed lines are contours
%of diffuse emission. Because of the on-board slew algorithms
%scans in that go near the anti-sun direction violate constraints.}
%\end{figure}

Although the PCA is not an imaging
instrument, the $ 1\deg$ collimators, when scanned in a pattern
across the galactic bulge, generate a response that can be fit to a model of
point sources plus diffuse emission associated with the center of the
galaxy. 
%The scan pattern is east-west or north-south on alternative
%observations about 3 days apart. 
%Figure 1 shows the scan pattern for one of the east-west scans.
The region scanned is a $16\deg \times 16\deg$ square with a diagonal 
along the galactic plane.
There is a gap  of coverage when the sun
is within $30\deg$ of the region and a smaller gap when the sun in the
Galactic anticenter. 
%A given source is passed over twice per observation.  
%If necessary the source intensity is allowed to vary.
At a scan rate of $6\deg$ per minute, a typical source is in the PCA field of
view only 20 seconds. However, RXTE operations require us to hold
the attitude briefly (3 minutes) between scans.
We chose our scan paths to stop on interesting sources.
%to the extent possible. 
New sources have been reported in I.A.U. Circulars 7103, 7120, and 7300.
%Markwardt et al. (1999), 
%Marshall \& Markwardt
%(1999a, 1999b), and Markwardt, Swank,  \& Marshall (1999).

%\section{Temporal Results}

\begin{figure}
% \plotfiddle{test.ps}{3.0in}{0.}{30}{30}{0}{0}
% \plotfiddle{test.ps}{3.0in}{0.}{50}{50}{0}{0}
\centerline{\epsffile{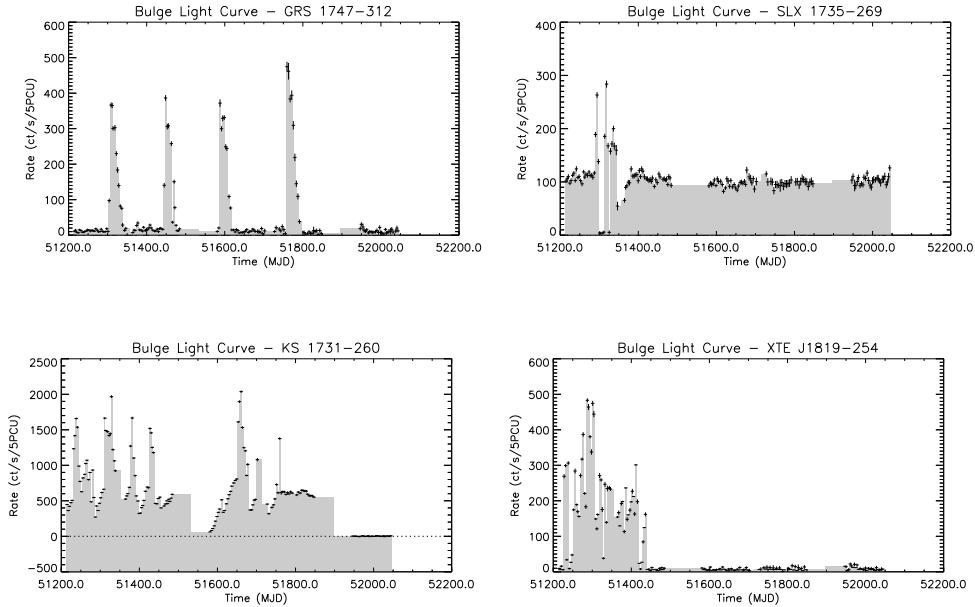}\ }
\caption{Light-curves of four of the monitored sources.
%Each data point is from a snap shot which represents a few minutes of data.
%Local galactic ridge background is fit to a model.
}
\end{figure}

There are several characteristic temporal behaviors. Examples are shown in 
Figure 1. 
There
are fast-rise/slow-fall transients (9 of them have been observed)
which take less than about 3 days to rise and about 3 weeks to decay.
In the first two
years of this monitoring program a number of sources had one such 
outburst (XTE~J1734--234, 
XTE~J1739--285, EXO~1745--248, SAX J1747--2853, and 
SAX~J1808.4--3658). Some, like
GRS~1747--312 in Terzan 6 and the Rapid Burster, have recurrence times,
(about 160 and 180 days, respectively)
which have allowed observation of multiple outbursts. 
XTE~J1739--285 is starting a second outburst after 580
days and all but XTE~J1734--234 (with peak flux of only  2.2 mCrab) 
have been seen to
have previous outbursts. 
Among the Atoll bursters, irregular cyclic
variations like the high and low states of 4U~1820--30
occur in 4 of the sources we have monitored. 
In some cases time-scales are
similar to the outbursts of the transients, but with a much greater
duty cycle, as in KS~1731--26.  Several of the sources seen in the
galactic bulge region have a high level of irregular variability. With
count rates of 10-200 counts s$^{-1}$ (scaled to 5 PCU) (about 1-20
mCrab), the large variations from observation to observation were very
significant for 9 additional sources resolved. The light curve for
SLX~1735--269 is one of the most peculiar. Occultation for 10 days
is an improbable interpretation. Dispersal of
an accretion disk may have occurred, with re-formation following
within a month. At least 5 of the sources had a flux with an average
steady level (and superposed variations) until the flux dropped precipitously
within a week.  Figure 1 shows the light-curve of 
XTE~J1819--254 $\equiv$ SAX~J1819.3--2525.
It was  identified as V4641 Sgr when it 
exhibited dramatic optical, X-ray (up to 12 Crab), and 
radio fireworks in September 1999.
These took place
entirely {\it between} the monitoring observations near MJD 51460.

Deeper observations have been carried out for some sources.
%although the observing times (3-25 ksec) were relatively short. 
No
pulsations have been seen similar to those of SAX~J1808.4--3658. For
that source 5 \% pulsations were seen when the source was only 10
mCrab (Wijnands et al. 2001). Limits of high frequency quasiperiodic
oscillations were typically a few percent. Observed power at
frequencies below 1 Hz was 10-30 \%. GRS~1747--312 in Terzan 6 
(in't Zand et al. 2000) and 
XTE~J1710--281 (Markwardt et al. 1999) are eclipsing. We do not
report on spectral information here. It is difficult to use the
spectra to separate black hole candidates from neutron star sources,
since both could have similar spectra in certain states. A burst 
with exponential decay strongly points to a neutron star
identification and the WFC has identified bursts from more than half of
the sources we have tracked.

\section{Discussion}

There have been several transients whose outburst characteristics are
very similar to those of SAX~J1808.4-3658. There have also been
additional outbursts by sources which are very much like Aql~X--1 and
4U~1608--52. The latter are about 3.5 kpc away
and their peak outburst rates of 500-1000 mCrab imply peak persistent
luminosities of $2-4 \times 10^{37} $ ergs s$^{-1}$. Thus they reach 0.1-0.2
of the Eddington limit. The less well known 
GRS~1747--312 and EXO~1745-248 are both at about 
7.6 kpc (in't Zand et al. 2000; Johnston, Verbunt, \& Hasinger 1995). 
Like SAX~J1808.4-3658, GRS~1747--312 has a peak
outburst luminosity 0.04 of the Eddington limit, while the peak
%$L_x/L_{Edd}$ 
ratio for an  outburst of EXO~1745-248 was 0.4.
As in't Zand (2000) has noted, the population is a mixture.
Distances are essential for determining whether some of these
transients have systematically less flux. 
%Bursts provide an estimate
%of distance if assumed to be a standard candle.  
A majority of
bursts may be standard candles within a factor of 2,  
but several sources
have appeared to be implausibly far if the bursts were assumed that
bright 
%The most extreme case was that of the ASCA observation of a
%source in M28 
(e.g. Gotthelf \& Kulkarni 1997). 

Binary periods are known for a handful of sources (Aql~X--1 19 hr;
GRS~1747--312 12 hr; AX~J1745.6-2901 8 hr;
SAX~J1808.4-3658 2 hr). The period of 
XTE~1710--281 is probably near 3 hr.  
In many of the binaries the secondary comes
into contact long before the minimum period of $\approx$ 80 minutes is
reached. Consistently, the outburst decay times
vary. King (2000) found a 4 day exponential time-scale in accord
with the short orbital period of SAX~J1808.4--3658. Others have linear
decays of several tens and even hundreds of days. The relation between 
the orbital period and the decay time-scale needs further study.

The average accretion rates from the companions to the disk that are
implied for the recurrent transients are in the range expected to be
unstable; that is, they are low enough that the X-ray flux irradiation
combined with the viscous dissipation are not enough to keep the outer
part of the disk ionized. Some are near the instability line of
$L_{av}$ versus $P_{orb}$ estimated by van Paradijs (1996).

While typical transients have a fast rise, a peak of several days, and
a manifest decay over many days, we have observed several examples of
sources which have flat topped light-curves (with high root mean
squared variability), and then fast decay, within a few
days. V4641~Sgr is the
most dramatic example of how the flux can erratically increase just
before the X-ray flux declines. 
%It suggests that if there is a disk, or
%discrete rings of accretion, they are 
Possibly a disk-like accretion reservoir is 
blown away, into the radio jet
in the case of V4641~Sgr, or completely
emptied and not replenished. This source is a black hole
system, but bursts from two of the sources which suddenly
die away suggest they are neutron stars. Spectra from the decay 
of GRS~1758--258 show the source declining rather than being swamped
(Smith et al. 2001).

For very low average accretion rates, neutron star sources are 
expected to be transient, 
with hardly more than  thermal radiation from the neutron
star (Rutledge et al. 1999), at about  $10^{33}$ ergs s$^{-1}$.
For more than 2 years  we have observed about 15 sources  
persisting at 0.5-80 mCrab,
some varying strongly, others steady within errors.
%the limits of the statistics and the systematics. 
The limiting intensity
for the PCA scans is about 0.5 mCrab, $1
\times 10^{35}$ ergs s$^{-1}$ at a galactic center distance of 8.5 kpc, 
or, for accretion, $1.5 \times 10^{-11}$ M$_{\sun}$ yr$^{-1}$, at which even
short period systems would be transient. 
At least one 
source, 2S~1803--245$\equiv$ XTE~J1806--246, is a transient in ``quiescence''
and more may be. 
%The population is arguably  slowly on its way to dropping out of the 
%persistent source luminosity distribution, as the flux goes lower and 
%cannot support ionization. 
%But at least some have variable quiescent flux 
%and may still be accreting then. 
If the magnetic field is as low as $5 \times 10^7$ gauss, implied accretion 
rates are too high to be inhibited by propeller effects, 
even for fast rotating neutron stars.
%, so the flux could still be from accretion.
Some of the transients 
may well recede to the levels of thermal emission, 
but a set of them may keep accreting. 
%To distinguish possible reasons requires 
%further information.

%\section{Conclusions}

%Observations are finding representatives of accretion rates varying
%from Eddington down to the low limits ($10^{-11}$ M$_{\sun}$yr$^{-1}$)
%for binary periods near the minimum period.

%Persistent sources operate of the verge of instability. No explanation
%has been advanced.

%Sources can be stably ``on'' at a low level for a long time and decay
%in 1-3 days, sometimes with a precursor short flaring.

%A significant number of transients discovered with RXTE and BeppoSax
%remain candidates for being short period binaries similar to
%SAX~J1808.6-3658, having low peak luminosity (0.03 Eddington)
%reflective of small disks. We hope to continue to examine them when
%they are active and to further define the characteristics of the
%population.

%It will take the imaging experiments Chandra, XMM, and in some case
%BeppoSAX to test whether, when the average accretion is very low in
%between outbursts, the transient's accretion turns off.

\section{References}

Chakrabarty, D., \& Morgan, E. H. 1998, Nature, 394,346\\
%Cui, W. et al. 
Smith, D. M., et al. 2001, \apj, in press, astro-ph/0103381\\
Gotthelf, E. V., \& Kulkarni, S. R. 1997, \apj, 490, L490\\
%Hasinger, G.  \& van der Klis, M. 1989, \aap, 225, 79\\
Heise, J., et al. 1999, Astro. Lett. and Communications, 38, 297\\
in't Zand, J. J. M.  2000, in 4th Integral workshop, 
Exploring the Gamma-Ray Universe,
in press\\
in't Zand, J. J. M., et al. 2000, \aap, 355, 145\\
Johnston, H. M., Verbunt, F., \& Hasinger, G. 1995, \aap, 298, L21\\
King, A.R. 2000, \mnras, 315, L33\\
%Markwardt, C. B. et al. 1999, IAUC 7300\\
%Marshall,F. E.  \& Markwardt, C. B. 1999b, IAUC 7133\\
%Marshall, F. E. \& Markwardt, C. B. 1999a, IAUC 7103\\
%Markwardt, F. E., Swank, J. H. \& Marshall, F. E. 1999, IAUC 7120\\
Markwardt, C. B., et al. 1999, \baas, 194, 8406\\
Rutledge, R. E., et al. 1999, \apj, 514, 945\\
Sidoli, L., et al. 1998, \aap, 336, L81\\
van Paradijs, J. 1996, \apj, 464, L139\\
Wijnands, R., et al. 2001, \apj, submitted, astro-ph/0105446\\
 
\end{document}